\documentclass[preprint,showpacs,amsmath,amssymb]{revtex4-2}

\usepackage{amsmath}
\usepackage{graphicx}
\usepackage{amsfonts}
\usepackage{dcolumn}
\usepackage{bm}
\usepackage{color}

\newcommand \f{\frac}

\begin{document}

\title{Mechanism of the Nonequilibrium Phase Transition in Self-Propelled Particles with Alignment}

\author{Ruizhe Yan}
\affiliation{Postgraduate training base Alliance of Wenzhou Medical University, Wenzhou, Zhejiang 325000, China}
\affiliation{Center for Theoretical Interdisciplinary Sciences, Wenzhou Institute, University of Chinese Academy of Sciences, Wenzhou, Zhejiang 325001, China}
\author{Jie Su}
\thanks{E-mail: sj0410@ucas.ac.cn}
\affiliation{Center for Theoretical Interdisciplinary Sciences, Wenzhou Institute, University of Chinese Academy of Sciences, Wenzhou, Zhejiang 325001, China}
\author{Jin Wang}
\thanks{E-mail: jin.wang.1@stonybrook.edu}
\affiliation{Postgraduate training base Alliance of Wenzhou Medical University, Wenzhou, Zhejiang 325000, China}
\affiliation{Center for Theoretical Interdisciplinary Sciences, Wenzhou Institute, University of Chinese Academy of Sciences, Wenzhou, Zhejiang 325001, China}
\affiliation{Department of Chemistry and of Physics and Astronomy, State University of New York of Stony Brook, Stony Brook, NY 11794, USA}
\date{\today}

\begin{abstract}
Self-propelled particles with alignment, displaying ordered collective motions such as swarming, can be investigated by the well-known Vicsek model. However, challenges still remain regarding the nature of the associated phase transition. Here, we use the landscape-flux approach combined with the coarse-grained mapping method to reveal the underlying mechanism of the continuous or discontinuous order-disorder nonequilibrium phase transition in Vicsek model systems featuring diverse noise characteristics. It is found that the nonequilibrium flux inside the landscape in the density-alignment degree phase space always rotates counterclockwise, and tends to delocalize or destabilize the point attractor states, providing the dynamical driving force for altering the landscape shape and the system state. Furthermore, the variations in the averaged flux and entropy production rate exhibit pronounced differences across various noise types. This not only helps to reveal the dynamical and thermodynamical mechanisms of the order-disorder transition but also offers a useful tool to recognize the continuity of the transition. Our findings present a novel perspective for exploring nonequilibrium phase transition behaviors and other collective motions in various complex systems.
\end{abstract}

\maketitle
\emph{Introduction}.--Self-propelled particles, widely existing as both natural objects and artificial swimmers \cite{a14,wioland2013confinement,a2,a11,a7,malescio2003stripe,jagla1998phase,dotera2014mosaic,a13,zhang2016natural,a12}, have the ability to execute purposeful systematic movement by taking in and dissipating energy, which causes the system to be far from the equilibrium and to exhibit many novel dynamical behaviors different from those in the equilibrium system. For instance, self-propelled particles can spontaneously organize into chains \cite{a14,a7}, vortexes \cite{wioland2013confinement,a2,bricard2015emergent}, clusters \cite{a11,a7,a12,buttinoni2013dynamical}, and notably, exhibit ordered collective motion and swarming behaviors \cite{a11,a7,reynolds1987flocks,vicsek1995novel}. Among diverse theoretical models of self-propelled particles, the Vicsek model with an alignment term \cite{vicsek1995novel} can successfully reproduce the ordered collective motion observed in nature like bird flocks and bacterial colonies. Notably, the Vicsek model system undergoes a phase transition from an ordered to a disordered state dependent on the decreasing particle density or the increasing noise intensity. Inspired by this, a series of studies \cite{a9,chepizhko2013optimal,liebchen2017collective,nagai2015collective,holubec2021finite,morin2015collective,costanzo2019milling,gulich2018temporal,roy2019effect,wu2021pattern,lu2022improved,you2023modified} have delved into the Vicsek model and its variation forms, such as introducing density-dependent particle velocity \cite{a9}, obstacles \cite{chepizhko2013optimal}, chirality \cite{liebchen2017collective}, memory effect \cite{nagai2015collective}, delayed effect \cite{holubec2021finite}, etc \cite{morin2015collective,costanzo2019milling,gulich2018temporal,roy2019effect,wu2021pattern,lu2022improved,you2023modified}, so that many novel collective motions emerge spontaneously. For example, when the Vicsek particle (VP) velocity is correlated with its local density \cite{a9}, abundant dynamical patterns can be observed, including moving clumps, active lanes, and asters. These findings underscore the practical potential of VP systems across various fields, including robotics \cite{lei2023exploring}, drone technology \cite{liu2021hierarchical} and so on.

For the original Vicsek model \cite{vicsek1995novel}, the phase transition from the ordered collective motion of VPs to the disordered one was considered continuous, which is proved by the smooth transition of the average normalized velocity dependence on the noise intensity and their scaling behavior \cite{vicsek1995novel}. Interestingly, when the intrinsic noise (also called the scalar or angular noise) is replaced by the extrinsic noise (the vectorial noise), not only do the travelling bands in the coexisting phase before the phase transition point emerge more easily, but also the variation in the noise-dependent average normalized velocity becomes sharper, both indicating that the order-disorder phase transition changes to a discontinuous one \cite{gregoire2004onset,nagy2007new,chate2008collective,cavagna2021vicsek}. The thermodynamical entropy of the system was estimated by investigating the compression-based entropy of the VP system under intrinsic or extrinsic noise \cite{cavagna2021vicsek}. However, only position (density) information was used to analyze the phase transition \cite{cavagna2021vicsek}, while the velocity information was ignored. Motivated by this, it is very necessary to establish a framework based on both position (density) and velocity information of VPs, which not only can present clearly and visually the order-disorder phase transition behaviors for different noise types, but also allow us to reveal the underlying dynamical and thermodynamical origins of the associated nonequilibrium phase transition.

	To address this, we introduce the nonequilibrium landscape and flux approach \cite{wang2008potential}, which is widely used in various fields, including cell fate decision making \cite{zhu2024uncovering}, cancer research \cite{li2015quantifying}, neural networks \cite{yan2013nonequilibrium}, ecological systems \cite{xu2021unifying}, among others \cite{li2014landscape,fang2019nonequilibrium,wang2015landscape,wu2013landscape}. In this approach, the driving forces for the nonequilibrium dynamics of the system are determined by both the landscape and rotational flux, analogous to the driving force of charged particle in electric field (potential gradient force) and magnetic field (Lorentz force). In this work, by using this approach combined with the coarse-grained mapping method \cite{su2024dynamical}, we establish and visualize the nonequilibrium potential and flux landscape of VP systems in the phase space of the local density and the local alignment degree, which reflects global states of the system as well as changes in dynamical and thermodynamical characteristics. As the noise intensity increases, the landscape changes continuously/discontinuously from one potential basin located at a large local alignment degree (the ordered flocking phase) to two basins (the coexisting phase consisting of travelling bands), and finally to one basin located at a small local alignment degree (the disordered phase) under intrinsic/extrinsic noise, respectively. Furthermore, the nonequilibrium flux inside the potential always rotates counterclockwise, providing the dynamical driving force for the shape changes of the landscape. Moreover, the obtained averaged flux and entropy production rate (EPR) exhibit distinct variation behaviors near the transition threshold, which not only can be utilized to reveal the dynamical and thermodynamical origins of the phase transition, but also serve as early warning signals of the transition. Notably, the variation behavior near the transition threshold differs in intensity for different noise types, offering an effective tool to recognize the continuity of the order-disorder transition. Additionally, forward and backward optimal transition paths between the two phases in the coexisting phase do not overlap with each other, indicating that the time-reversal symmetry of the VP system is broken.

\emph{Model}.--We consider a quasi two-dimensional system with size $W$ under periodic boundary conditions, including $N$ VPs. The motion of the $i$-th particle located at $\mathbf x_i$ obeys the following equation:

\begin{equation}
\mathbf x_i(t+\Delta t)=\mathbf x_i(t)+\mathbf v_i(t+\Delta t).\label{eq:translation1}
\end{equation}

\noindent Herein, $\mathbf v_i(t)\equiv v_0(\cos\theta_i(t),\sin\theta_i(t))$, where $v_0$ and $\theta_i$ denote respectively the amplitude and direction of $\mathbf v_i$. The evolution rule of $\theta_i$ for the VP under intrinsic noise \cite{vicsek1995novel} is given as

\begin{equation}
\theta_i(t+\Delta t)=\arg\Bigg(\sum_{r_{ij}<R}\mathbf v_j(t)\Bigg)+\eta\xi_i(t).\label{eq:rotation1}
\end{equation}

\noindent Here, the first term represents the alignment effect of the VP, i.e., $\theta_i$ always tends to orient towards the average direction of its neighbors, whose distance ($r_{ij}$) from the the given particle is smaller than the interaction radius $R$. The second term denotes the intrinsic noise, with $\eta$ the noise intensity and $\xi_i$ the white noise distributed uniformly ranging from $-\pi$ to $\pi$. For the situation of extrinsic noise, the evolution of $\theta_i$ should be replaced by \cite{cavagna2021vicsek}:

\begin{equation}
\theta_i(t+\Delta t)=\arg\Bigg(\sum_{r_{ij}<R}\mathbf v_j(t)+\eta\boldsymbol\xi_i(t)\Bigg),\label{eq:rotation1}
\end{equation}

\noindent where $\boldsymbol\xi_i$ is a unit vector with a random direction, i.e., $\boldsymbol\xi_i=(\cos(\zeta_i),\sin(\zeta_i))$ with $\zeta_i$ a random number selected from $[0,2\pi]$. In the rest of this paper, we fix $R=1$, $v_0=0.5$, $W=100$, $N=20000$, $\Delta t=1$, and average density $\rho_0=N/W^2=2$. We first run each simulation for $100000$ time steps to ensure the system reaches the steady state. Then, we perform the simulation for another $200000$ time steps in the steady state for sampling and analysis.

\begin{figure}
\centering
\includegraphics[width=1.0\columnwidth]{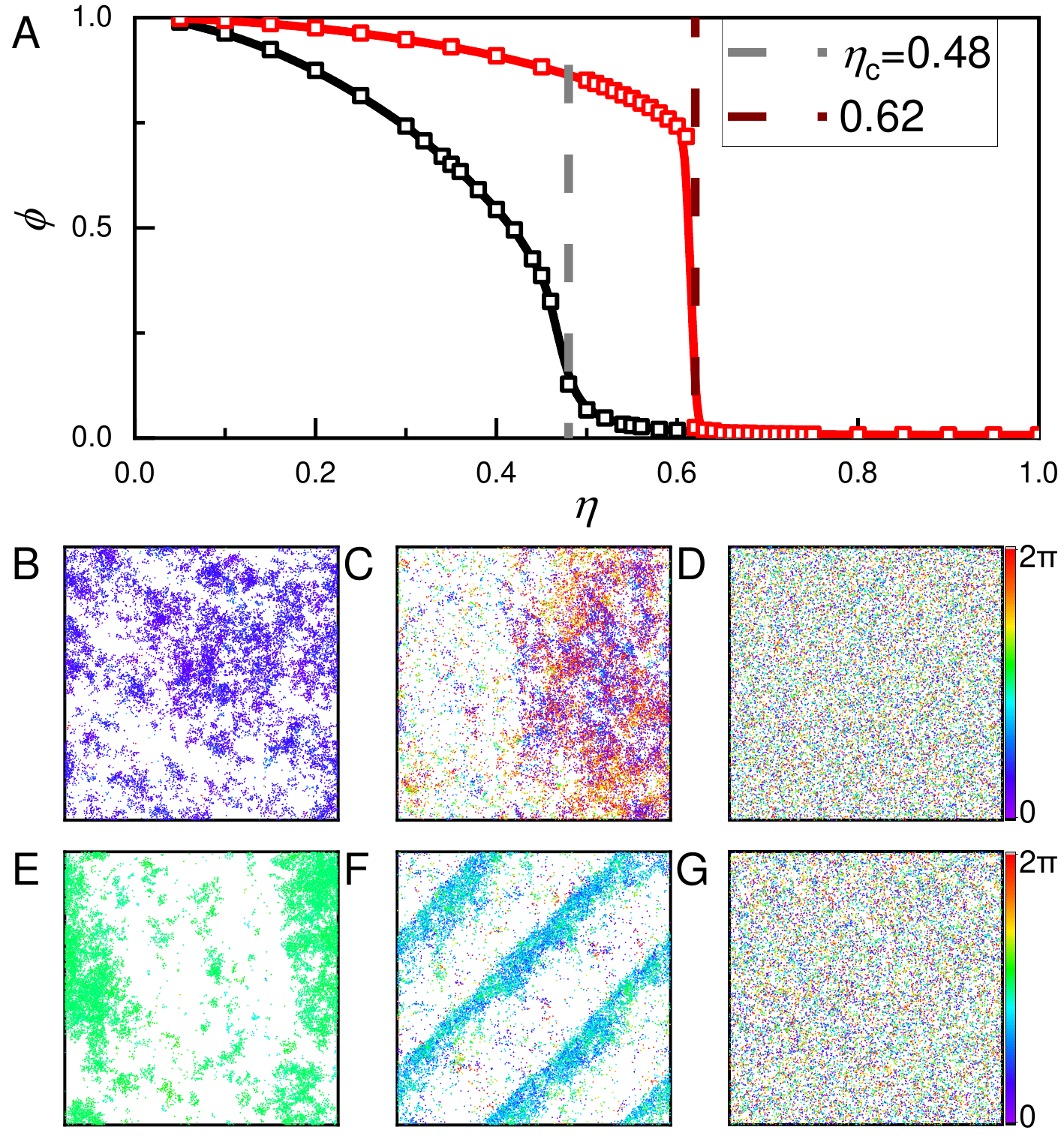}
\caption{(A) $\phi$ vs $\eta$ for the VP system under intrinsic noise (black symbols and line) and extrinsic noise (red symbols and line), with dashed lines the critical transition points $\eta_{\rm c}$. (B-G) Typical snapshots in the steady states for different phases with $\eta=0.1$ (B), $0.4$ (C) and $0.9$ (D) respectively under intrinsic noise as well as with $\eta=0.1$ (E), $0.6$ (F) and $0.8$ (G) respectively under extrinsic noise. (B) and (E) represent the ordered flocking phases. (C) and (F) denote the coexisting phases with travelling bands. (D) and (G) show the disordered phases.}
\label{fig:result1}
\end{figure}

\emph{Order-disorder phase transition for different noise}.--By gradually increasing the noise strength $\eta$ from $0.05$ to $1$, we observe transitions of the VP system from an ordered phase to a disordered one. To measure the order degree of the VP system, we first introduce the average normalized velocity $\phi=|\sum^N_{i=1}\mathbf v_i|/(Nv_0)$ (Figure~\ref{fig:result1}A), where $\phi=1$ represents VPs moving in the same orientation so that the system is in the ordered phase, while $\phi$ being close to $0$ implies VPs moving in random directions such that the system is disordered. For intrinsic noise (the black symbols and line in Figure~\ref{fig:result1}A), it is observed that as the noise intensity $\eta$ increases, $\phi$ decreases smoothly from almost $1$ to $0$, indicating that the VP system changes continuously from the ordered phase to the disordered one. For small noise such as $\eta=0.1$ as shown in Figure~\ref{fig:result1}B, it is found that VPs gather into clusters and move in similar directions, so that the VP system forms the flocking phase with $\phi$ close to $1$. For larger noise such as $\eta=0.4$ (Figure~\ref{fig:result1}C), many VPs gather and move collectively in one direction, further forming a travelling band of the dense phase, while the other particles move randomly in the dilute phase. For noise large enough such as $\eta=0.9$ (Figure~\ref{fig:result1}D), the motion of VPs becomes almost random, and the system becomes disordered. When the intrinsic noise is replaced by extrinsic noise (red symbols and line in Figure~\ref{fig:result1}A), $\phi$ still decreases from almost $1$ to $0$. However, the transition is much sharper than that under intrinsic noise. For small and large enough noise, the VP system also forms the ordered flocking phase (Figure~\ref{fig:result1}E) and disordered phase (Figure~\ref{fig:result1}G), respectively. Interestingly, when noise changes from intrinsic to extrinsic, the coexisting phase that appears at moderate noise levels shows a clearer travelling band.

\emph{Coarse-grained mapping method with landscape and flux}.--In order to reveal the origin of the order-disorder phase transition in the VP system under intrinsic or extrinsic noise, we use the coarse-grained mapping method \cite{su2024dynamical} and the landscape-flux approach \cite{wang2008potential,fang2019nonequilibrium}. By applying the coarse-grained mapping method, the information in the real space obtained from Eq.(\ref{eq:translation1}-\ref{eq:rotation1}) can be mapped into the phase space based on the local density $\rho$ and local alignment degree $q$ (more details and definitions can be found in the supplemental information, SI).

On this basis, the evolution of the system in the $\rho-q$ phase space can be obtained, which can be written as:

\begin{equation}
\dot{\mathbf X}=\mathbf F(\rho,q)+\boldsymbol\xi(\rho,q),
\end{equation}

\noindent where $\mathbf X=(\rho,q)$. Also, the steady-state probability distribution $P(\rho,q)$, the driving force $\mathbf F(\rho,q)$ and the diffusion coefficient matrix $\mathbf D(\rho,q)$ can be calculated from the simulation data. According to the landscape-flux approach, the driving force for the nonequilibrium dynamics is determined by both the nonequilibrium potential landscape and the nonequilibrium flux of the VP system in the $\rho-q$ phase space being quantified as (details can be found in the SI):

\begin{equation}
U(\rho,q)=-\ln P(\rho,q),
\label{eq:Uneq}
\end{equation}
\begin{equation}
\mathbf J(\rho,q)=\mathbf F(\rho,q)P(\rho,q)-\bm\nabla\cdot[\mathbf D(\rho,q)P(\rho,q)].
\label{eq:Flux}
\end{equation}

\emph{Landscape and flux in the phase space}.--Firstly, we quantify the potential landscape in the $\rho-q$ phase space (see Figure S4 for typical potential) and observe that the number of potential basins changes from $1$ to $2$ and finally back to $1$ for both intrinsic and extrinsic noise, in agreement with the phase transition behavior exhibited in the VP system (Figure~\ref{fig:result1}). More interestingly, the landscape changes continuously when the system undergoes intrinsic noise, but it changes discontinuously for extrinsic noise, demonstrating that the order-disorder phase transition is more discontinuous for extrinsic noise.

\begin{figure}
\centering
\includegraphics[width=1.0\columnwidth]{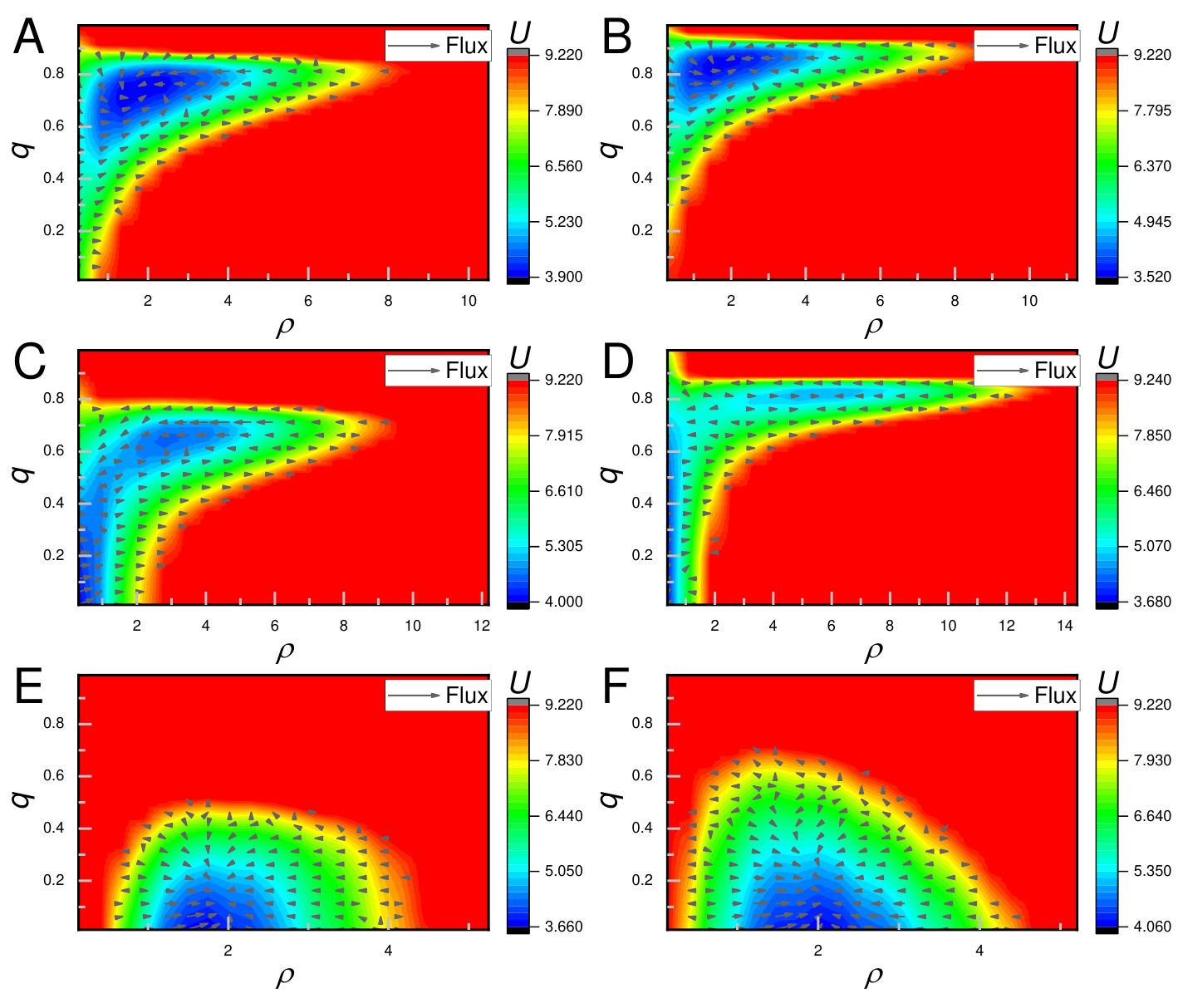}
\caption{The nonequilibrium potential and flux landscape in the $\rho-q$ phase space of the VP system under intrinsic (A, C, E) and extrinsic (B, D, F) noise. The noise intensity is $\eta=0.3$ (A), $0.4$ (C), $0.5$ (B), $0.6$ (D, E), and $0.62$ (F), respectively. The flux is represented by the black arrows, and the nonequilibrium potential is presented by the colored background.}
\label{fig:result2}
\end{figure}

We then focus on the dynamical nature of the VP system by presenting the nonequilibrium flux inside the landscape in the $\rho-q$ phase space (Figure~\ref{fig:result2}). The non-zero nonequilibrium flux demonstrates that the detailed balance of the VP system is broken. For the ordered flocking phase in the VP system, both under intrinsic and extrinsic noise (Figure~\ref{fig:result2}A and B), there is a counterclockwise rotating flow around the potential well. Different from the gradient force resulting from the nonequilibrium potential, which always attracts the system down into the potential basin, the curl flux favors global movements so that the system can not be localized in the basin any more, providing the dynamical driving force to create the coexisting phase with the ordered dense phase and the disordered dilute phase. For the coexisting phase (Figure~\ref{fig:result2}C and D), it is observed that the flux continues to rotate counterclockwise between the ordered dense phase and the disordered dilute phase, demonstrating that the nonequilibrium flux serves as the dynamical driving force for the order-disorder phase separation. When noise becomes sufficiently large (Figure~\ref{fig:result2}E and F), the rotation of the nonequilibrium flux inside the single potential basin becomes weak and indistinct due to the large noise.

\begin{figure}
\centering
\includegraphics[width=1.0\columnwidth]{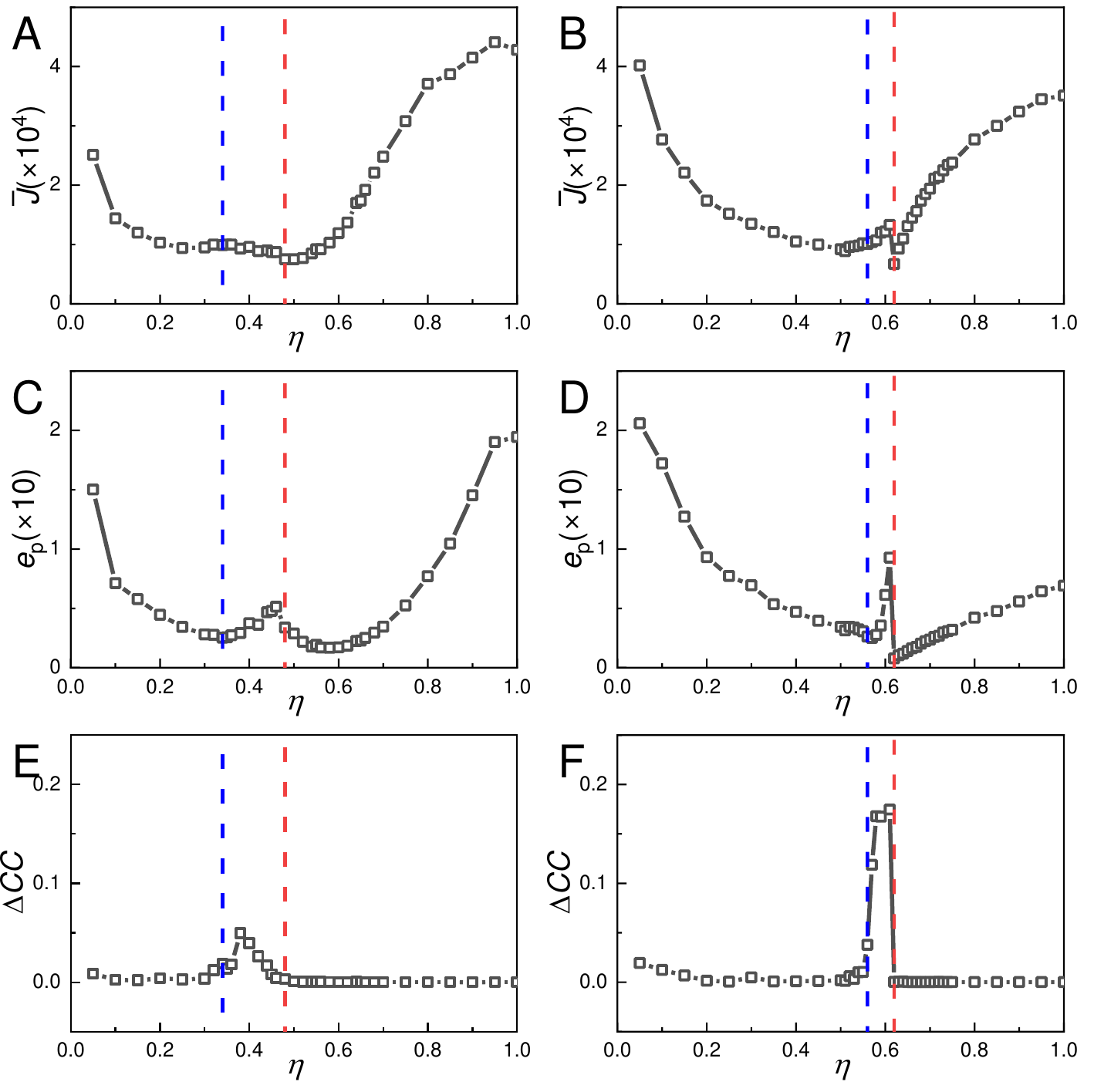}
\caption{Dynamical and thermodynamical mechanisms of the phase transition in the VP system. Plots of $\bar J$ (A, B), $e_{\rm p}$ (C, D) and $\Delta CC$ (E, F) versus the noise strength $\eta$ under intrinsic (A, C, E) and extrinsic (B, D, F) noise. The blue/red dashed lines ($\eta_{\rm b}$/$\eta_{\rm c}$) represent the phase boundaries between the coexisting phase and the ordered phase/the disordered phase.}
\label{fig:result3}
\end{figure}

\emph{Dynamical and thermodynamical origins of nonequilibrium phase transition}.--We now explore the dynamical and thermodynamical origins of the phase transition. To reveal the dynamical origin, we quantitatively calculate the contribution of the nonequilibrium flux in the phase transition process, i.e., the dependence of the averaged flux $\bar J$ on $\eta$ as shown in Figure~\ref{fig:result3}A and B. For the VP system under intrinsic noise (Figure~\ref{fig:result3}A), as $\eta$ increases, $\bar J$ first decreases in the region of the ordered flocking phase, remains relatively unchanged with fluctuations in the coexisting phase, and finally increases in the disordered phase. This means that the slope of $\bar J$ changes from $0$ to a positive value when $\eta$ crosses the order-disorder phase transition threshold $\eta_{\rm c}$. In contrast to the case of intrinsic noise, it is observed that $\bar J$ initially increases but later decreases in the coexisting phase when noise is extrinsic (Figure~\ref{fig:result3}B), resulting in a peak appearing just before the phase transition threshold. In other words, the slope of $\bar J$ changes from a negative value to a positive one as $\eta$ exceeds $\eta_{\rm c}$. Physically, the flux as one of the main driving forces for the nonequilibrium system dynamics prefers to delocalize the states due to its rotational nature rather than localizing or converging states by potential gradient force. Thus the variation of the flux favors destabilizing the current state and generating a new state. The variations in the slopes of $\bar J$ near the phase transition threshold imply that the nonequilibrium flux is the dynamical origin of the order-disorder phase transition. More interestingly, we notice that compared to the situation under intrinsic noise, the variation of $\bar J$ is more sudden and sharp under extrinsic noise, suggesting that the order-disorder phase transition is more discontinuous under extrinsic noise.

Now we focus on the thermodynamics of the VP system under different noise types. To quantify the detailed balance breaking and time-reversal symmetry violation in the $\rho-q$ phase space of the VP system, we introduce the entropy production rate $e_{\rm p}$ of the system \cite{su2024dynamical}:

\begin{equation}
e_{\rm p}=\iint\mathrm d\rho\mathrm dq\f{\mathbf J(\rho,q)\mathbf D^{-1}(\rho,q)\mathbf J^{\rm T}(\rho,q)}{P(\rho,q)}.
\end{equation}

\noindent As shown in Figure~\ref{fig:result3}C and D, for both intrinsic and extrinsic noise, $e_{\rm p}$ decreases in the ordered flocking phase, but increases in the coexisting phase and the disordered phase, further exhibiting a peak at the boundary between these two phases. Since entropy production rate measures the thermodynamic cost of the action of the flux destabilizing the current state and generating a new state, the emergence of this peak near the boundary between the ordered and disordered phases provides an indicator for the thermodynamical origin of the nonequilibrium phase transition in the VP system. Furthermore, it is observed that the peak of $e_{\rm p}$ in extrinsic noise is sharper than that in intrinsic noise, demonstrating again that the phase transition is more discontinuous under extrinsic noise compared to that under intrinsic noise.

Finally, we introduce the average difference $\Delta CC$ between the two-point cross correlations forward and backward in time (more details and definitions can be found in SI), as plotted in Figure~\ref{fig:result3}E and F. It is observed that $\Delta CC$ shows a peak in the phase separation region for both intrinsic and extrinsic noise, illustrating that the time irreversibility of the VP system reaches its maximum during the phase transition process. The sharper peak of $\Delta CC$ under extrinsic noise indicates again the more discontinuous nature of the phase transition with a clearer travelling band emergence under extrinsic noise conditions. More interestingly, $\bar J$, $e_{\rm p}$ and $\Delta CC$ all show significant variations prior to reaching the critical transition point $\eta_{\rm c}$. Along with the peak appearance, this provides early warning signals for the order-disorder phase transition in the VP system with both intrinsic and extrinsic noise.

\begin{figure}
\centering
\includegraphics[width=1.0\columnwidth]{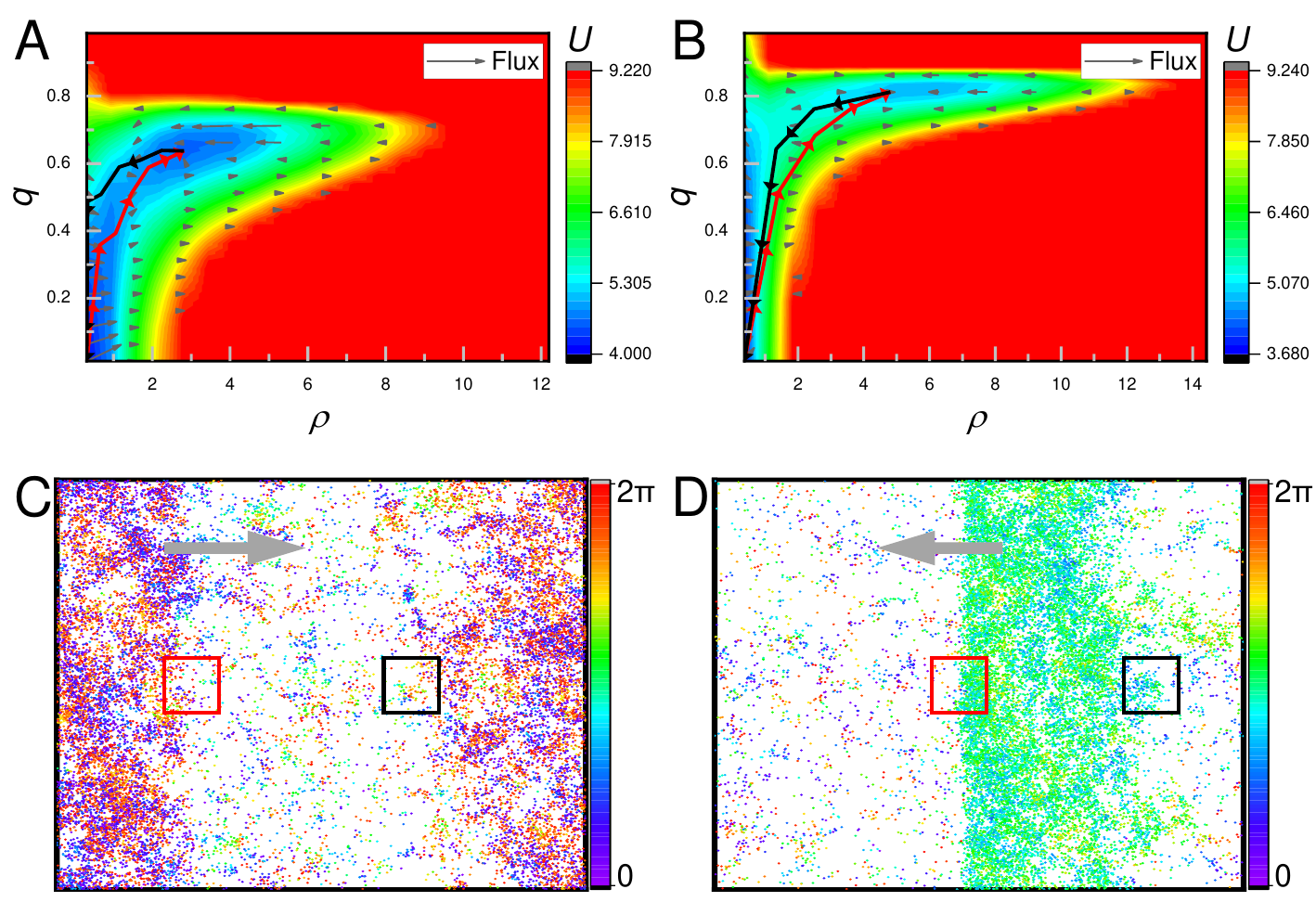}
\caption{Optimal transition paths between dense and dilute phases in the coexisting phase of the VP system. A, B show the backward (black lines) and forward (red lines) optimal transition paths between two phases under intrinsic ($\eta=0.4$) and extrinsic ($\eta=0.6$) noise respectively. C, D display typical snapshots in the steady states corresponding to A and B respectively. The arrows represent the moving directions of the travelling bands.}
\label{fig:result4}
\end{figure}

\emph{Optimal transition paths in the coexisting phase}.--Furthermore, to unravel the fundamental mechanisms of the dynamical process between the dense and dilute phases in the coexisting phase, we employ the landscape and flux approach to analyze the optimal transition paths \cite{wang2010kinetic}. As depicted in Figure~\ref{fig:result4}A and B, it is evident that for both intrinsic and extrinsic noise, the backward optimal transition paths (black lines) originating from the dense phase initially exhibit a slow decrease in $q$ but a rapid decrease in $\rho$, followed by a sharp decline in $q$ until reaching the dilute phase. Conversely, the forward optimal transition paths (red lines) switching from the dilute phase to the dense phase involve simultaneous increases in both $\rho$ and $q$. Notably, the forward and backward optimal transition paths do not overlap with each other, indicating that the time-reversal symmetry is broken. These non-overlapped paths result from the combined effect of the potential and the curl flux. Next, we focus on the dynamical behaviors of the travelling band in the coexisting phase. As shown in Figure~\ref{fig:result4}C and D, it is found that for both intrinsic and extrinsic noise, compared to the region ahead of the band (red boxes) with VPs originating from the disordered phase, the area behind the band (black boxes) primarily comprises VPs that have just exited the travelling band. Consequently, at similar $\rho$ values, the black boxes exhibit larger $q$ values than the red boxes, which aligns with the observation that the backward optimal transition path (black lines in Figure~\ref{fig:result4}A and B) consistently lies above the forward path (red lines).

\emph{Conclusion}.--In summary, by using the landscape-flux approach combined with the coarse-grained mapping method in the VP system, we quantified the nonequilibrium potential and flux field in the $\rho-q$ phase space for both intrinsic and extrinsic noise. Our results not only reflect the global states of the system changing continuously/discontinuously from the ordered flocking phase to the disordered phase under intrinsic/extrinsic noise, but also reveal the dynamical and thermodynamical mechanisms of the phase transition. Physically, due to its rotational nature, flux tends to delocalize or destabilize the point attractors while entropy  production rate provides the thermodynamic cost. The variation in slope or the emergence of a peak in the averaged flux or the EPR near the boundary between ordered and disordered phases supports that the nonequilibrium flux and the EPR are the dynamical and thermodynamical origins, respectively. The intensity of the slope variation under different noise types can be used to characterize the discontinuity of the order-disorder phase transition, and this variation can provide early warning signals of the transition. Furthermore, we plotted the optimal transition paths in the coexisting phase with the travelling band, demonstrating the breaking of time-reversal symmetry. Our findings not only provide a physical and visual way to reveal the underlying mechanism of the phase transition in the VP system, but also offer a new perspective to explore other collective behaviors in more complex nonequilibrium systems.

\emph{Acknowledgments}.--R.Y. and J.S. are supported by WIUCASQD2022012, NSFC12404237 and NSFC12234019.

\emph{Competing interests}.--The authors declare no competing interests.

\emph{Data availability}.--All study data are included in this article and/or the SI Appendix.

\emph{Author contributions}.--J.W. proposed and designed the research. J.S. provided the model and supervised the research. R.Y. performed the simulations and data analysis. All authors discussed the results and co-wrote the manuscript.


\begin{thebibliography}{44}
\expandafter\ifx\csname natexlab\endcsname\relax\def\natexlab#1{#1}\fi
\expandafter\ifx\csname bibnamefont\endcsname\relax
  \def\bibnamefont#1{#1}\fi
\expandafter\ifx\csname bibfnamefont\endcsname\relax
  \def\bibfnamefont#1{#1}\fi
\expandafter\ifx\csname citenamefont\endcsname\relax
  \def\citenamefont#1{#1}\fi
\expandafter\ifx\csname url\endcsname\relax
  \def\url#1{\texttt{#1}}\fi
\expandafter\ifx\csname urlprefix\endcsname\relax\def\urlprefix{URL }\fi
\providecommand{\bibinfo}[2]{#2}
\providecommand{\eprint}[2][]{\url{#2}}

\bibitem[{\citenamefont{Bechinger et~al.}(2016)\citenamefont{Bechinger,
  Di~Leonardo, L{\"o}wen, Reichhardt, Volpe, and Volpe}}]{a14}
\bibinfo{author}{\bibfnamefont{C.}~\bibnamefont{Bechinger}},
  \bibinfo{author}{\bibfnamefont{R.}~\bibnamefont{Di~Leonardo}},
  \bibinfo{author}{\bibfnamefont{H.}~\bibnamefont{L{\"o}wen}},
  \bibinfo{author}{\bibfnamefont{C.}~\bibnamefont{Reichhardt}},
  \bibinfo{author}{\bibfnamefont{G.}~\bibnamefont{Volpe}}, \bibnamefont{and}
  \bibinfo{author}{\bibfnamefont{G.}~\bibnamefont{Volpe}},
  \bibinfo{journal}{Rev. Mod. Phys.} \textbf{\bibinfo{volume}{88}},
  \bibinfo{pages}{045006} (\bibinfo{year}{2016}).

\bibitem[{\citenamefont{Wioland et~al.}(2013)\citenamefont{Wioland, Woodhouse,
  Dunkel, Kessler, and Goldstein}}]{wioland2013confinement}
\bibinfo{author}{\bibfnamefont{H.}~\bibnamefont{Wioland}},
  \bibinfo{author}{\bibfnamefont{F.~G.} \bibnamefont{Woodhouse}},
  \bibinfo{author}{\bibfnamefont{J.}~\bibnamefont{Dunkel}},
  \bibinfo{author}{\bibfnamefont{J.~O.} \bibnamefont{Kessler}},
  \bibnamefont{and} \bibinfo{author}{\bibfnamefont{R.~E.}
  \bibnamefont{Goldstein}}, \bibinfo{journal}{Physical review letters}
  \textbf{\bibinfo{volume}{110}}, \bibinfo{pages}{268102}
  (\bibinfo{year}{2013}).

\bibitem[{\citenamefont{Sumino et~al.}(2012)\citenamefont{Sumino, Nagai,
  Shitaka, Tanaka, Yoshikawa, Chat{\'e}, and Oiwa}}]{a2}
\bibinfo{author}{\bibfnamefont{Y.}~\bibnamefont{Sumino}},
  \bibinfo{author}{\bibfnamefont{K.~H.} \bibnamefont{Nagai}},
  \bibinfo{author}{\bibfnamefont{Y.}~\bibnamefont{Shitaka}},
  \bibinfo{author}{\bibfnamefont{D.}~\bibnamefont{Tanaka}},
  \bibinfo{author}{\bibfnamefont{K.}~\bibnamefont{Yoshikawa}},
  \bibinfo{author}{\bibfnamefont{H.}~\bibnamefont{Chat{\'e}}},
  \bibnamefont{and} \bibinfo{author}{\bibfnamefont{K.}~\bibnamefont{Oiwa}},
  \bibinfo{journal}{Nature} \textbf{\bibinfo{volume}{483}},
  \bibinfo{pages}{448} (\bibinfo{year}{2012}).

\bibitem[{\citenamefont{Karani et~al.}(2019)\citenamefont{Karani, Pradillo, and
  Vlahovska}}]{a11}
\bibinfo{author}{\bibfnamefont{H.}~\bibnamefont{Karani}},
  \bibinfo{author}{\bibfnamefont{G.~E.} \bibnamefont{Pradillo}},
  \bibnamefont{and} \bibinfo{author}{\bibfnamefont{P.~M.}
  \bibnamefont{Vlahovska}}, \bibinfo{journal}{Phys. Rev. Lett.}
  \textbf{\bibinfo{volume}{123}}, \bibinfo{pages}{208002}
  (\bibinfo{year}{2019}).

\bibitem[{\citenamefont{Yan et~al.}(2016)\citenamefont{Yan, Han, Zhang, Xu,
  Luijten, and Granick}}]{a7}
\bibinfo{author}{\bibfnamefont{J.}~\bibnamefont{Yan}},
  \bibinfo{author}{\bibfnamefont{M.}~\bibnamefont{Han}},
  \bibinfo{author}{\bibfnamefont{J.}~\bibnamefont{Zhang}},
  \bibinfo{author}{\bibfnamefont{C.}~\bibnamefont{Xu}},
  \bibinfo{author}{\bibfnamefont{E.}~\bibnamefont{Luijten}}, \bibnamefont{and}
  \bibinfo{author}{\bibfnamefont{S.}~\bibnamefont{Granick}},
  \bibinfo{journal}{Nat. Mater.} \textbf{\bibinfo{volume}{15}},
  \bibinfo{pages}{1095} (\bibinfo{year}{2016}).

\bibitem[{\citenamefont{Malescio and Pellicane}(2003)}]{malescio2003stripe}
\bibinfo{author}{\bibfnamefont{G.}~\bibnamefont{Malescio}} \bibnamefont{and}
  \bibinfo{author}{\bibfnamefont{G.}~\bibnamefont{Pellicane}},
  \bibinfo{journal}{Nature materials} \textbf{\bibinfo{volume}{2}},
  \bibinfo{pages}{97} (\bibinfo{year}{2003}).

\bibitem[{\citenamefont{Jagla}(1998)}]{jagla1998phase}
\bibinfo{author}{\bibfnamefont{E.}~\bibnamefont{Jagla}},
  \bibinfo{journal}{Physical Review E} \textbf{\bibinfo{volume}{58}},
  \bibinfo{pages}{1478} (\bibinfo{year}{1998}).

\bibitem[{\citenamefont{Dotera et~al.}(2014)\citenamefont{Dotera, Oshiro, and
  Ziherl}}]{dotera2014mosaic}
\bibinfo{author}{\bibfnamefont{T.}~\bibnamefont{Dotera}},
  \bibinfo{author}{\bibfnamefont{T.}~\bibnamefont{Oshiro}}, \bibnamefont{and}
  \bibinfo{author}{\bibfnamefont{P.}~\bibnamefont{Ziherl}},
  \bibinfo{journal}{Nature} \textbf{\bibinfo{volume}{506}},
  \bibinfo{pages}{208} (\bibinfo{year}{2014}).

\bibitem[{\citenamefont{Du et~al.}(2019)\citenamefont{Du, Jiang, and
  Hou}}]{a13}
\bibinfo{author}{\bibfnamefont{Y.}~\bibnamefont{Du}},
  \bibinfo{author}{\bibfnamefont{H.}~\bibnamefont{Jiang}}, \bibnamefont{and}
  \bibinfo{author}{\bibfnamefont{Z.}~\bibnamefont{Hou}}, \bibinfo{journal}{J.
  Chem. Phys.} \textbf{\bibinfo{volume}{151}}, \bibinfo{pages}{154904}
  (\bibinfo{year}{2019}).

\bibitem[{\citenamefont{Zhang and Granick}(2016)}]{zhang2016natural}
\bibinfo{author}{\bibfnamefont{J.}~\bibnamefont{Zhang}} \bibnamefont{and}
  \bibinfo{author}{\bibfnamefont{S.}~\bibnamefont{Granick}},
  \bibinfo{journal}{Faraday discussions} \textbf{\bibinfo{volume}{191}},
  \bibinfo{pages}{35} (\bibinfo{year}{2016}).

\bibitem[{\citenamefont{Gou et~al.}(2019)\citenamefont{Gou, Jiang, and
  Hou}}]{a12}
\bibinfo{author}{\bibfnamefont{Y.}~\bibnamefont{Gou}},
  \bibinfo{author}{\bibfnamefont{H.}~\bibnamefont{Jiang}}, \bibnamefont{and}
  \bibinfo{author}{\bibfnamefont{Z.}~\bibnamefont{Hou}}, \bibinfo{journal}{Soft
  Matter} \textbf{\bibinfo{volume}{15}}, \bibinfo{pages}{9104}
  (\bibinfo{year}{2019}).

\bibitem[{\citenamefont{Bricard et~al.}(2015)\citenamefont{Bricard, Caussin,
  Das, Savoie, Chikkadi, Shitara, Chepizhko, Peruani, Saintillan, and
  Bartolo}}]{bricard2015emergent}
\bibinfo{author}{\bibfnamefont{A.}~\bibnamefont{Bricard}},
  \bibinfo{author}{\bibfnamefont{J.-B.} \bibnamefont{Caussin}},
  \bibinfo{author}{\bibfnamefont{D.}~\bibnamefont{Das}},
  \bibinfo{author}{\bibfnamefont{C.}~\bibnamefont{Savoie}},
  \bibinfo{author}{\bibfnamefont{V.}~\bibnamefont{Chikkadi}},
  \bibinfo{author}{\bibfnamefont{K.}~\bibnamefont{Shitara}},
  \bibinfo{author}{\bibfnamefont{O.}~\bibnamefont{Chepizhko}},
  \bibinfo{author}{\bibfnamefont{F.}~\bibnamefont{Peruani}},
  \bibinfo{author}{\bibfnamefont{D.}~\bibnamefont{Saintillan}},
  \bibnamefont{and} \bibinfo{author}{\bibfnamefont{D.}~\bibnamefont{Bartolo}},
  \bibinfo{journal}{Nature communications} \textbf{\bibinfo{volume}{6}},
  \bibinfo{pages}{7470} (\bibinfo{year}{2015}).

\bibitem[{\citenamefont{Buttinoni et~al.}(2013)\citenamefont{Buttinoni,
  Bialk{\'e}, K{\"u}mmel, L{\"o}wen, Bechinger, and
  Speck}}]{buttinoni2013dynamical}
\bibinfo{author}{\bibfnamefont{I.}~\bibnamefont{Buttinoni}},
  \bibinfo{author}{\bibfnamefont{J.}~\bibnamefont{Bialk{\'e}}},
  \bibinfo{author}{\bibfnamefont{F.}~\bibnamefont{K{\"u}mmel}},
  \bibinfo{author}{\bibfnamefont{H.}~\bibnamefont{L{\"o}wen}},
  \bibinfo{author}{\bibfnamefont{C.}~\bibnamefont{Bechinger}},
  \bibnamefont{and} \bibinfo{author}{\bibfnamefont{T.}~\bibnamefont{Speck}},
  \bibinfo{journal}{Physical review letters} \textbf{\bibinfo{volume}{110}},
  \bibinfo{pages}{238301} (\bibinfo{year}{2013}).

\bibitem[{\citenamefont{Reynolds}(1987)}]{reynolds1987flocks}
\bibinfo{author}{\bibfnamefont{C.~W.} \bibnamefont{Reynolds}}, in
  \emph{\bibinfo{booktitle}{Proceedings of the 14th annual conference on
  Computer graphics and interactive techniques}} (\bibinfo{year}{1987}), pp.
  \bibinfo{pages}{25--34}.

\bibitem[{\citenamefont{Vicsek et~al.}(1995)\citenamefont{Vicsek, Czir{\'o}k,
  Ben-Jacob, Cohen, and Shochet}}]{vicsek1995novel}
\bibinfo{author}{\bibfnamefont{T.}~\bibnamefont{Vicsek}},
  \bibinfo{author}{\bibfnamefont{A.}~\bibnamefont{Czir{\'o}k}},
  \bibinfo{author}{\bibfnamefont{E.}~\bibnamefont{Ben-Jacob}},
  \bibinfo{author}{\bibfnamefont{I.}~\bibnamefont{Cohen}}, \bibnamefont{and}
  \bibinfo{author}{\bibfnamefont{O.}~\bibnamefont{Shochet}},
  \bibinfo{journal}{Physical review letters} \textbf{\bibinfo{volume}{75}},
  \bibinfo{pages}{1226} (\bibinfo{year}{1995}).

\bibitem[{\citenamefont{Farrell et~al.}(2012)\citenamefont{Farrell, Marchetti,
  Marenduzzo, and Tailleur}}]{a9}
\bibinfo{author}{\bibfnamefont{F.~D.~C.} \bibnamefont{Farrell}},
  \bibinfo{author}{\bibfnamefont{M.~C.} \bibnamefont{Marchetti}},
  \bibinfo{author}{\bibfnamefont{D.}~\bibnamefont{Marenduzzo}},
  \bibnamefont{and} \bibinfo{author}{\bibfnamefont{J.}~\bibnamefont{Tailleur}},
  \bibinfo{journal}{Phys. Rev. Lett.} \textbf{\bibinfo{volume}{108}},
  \bibinfo{pages}{248101} (\bibinfo{year}{2012}).

\bibitem[{\citenamefont{Chepizhko et~al.}(2013)\citenamefont{Chepizhko,
  Altmann, and Peruani}}]{chepizhko2013optimal}
\bibinfo{author}{\bibfnamefont{O.}~\bibnamefont{Chepizhko}},
  \bibinfo{author}{\bibfnamefont{E.~G.} \bibnamefont{Altmann}},
  \bibnamefont{and} \bibinfo{author}{\bibfnamefont{F.}~\bibnamefont{Peruani}},
  \bibinfo{journal}{Physical review letters} \textbf{\bibinfo{volume}{110}},
  \bibinfo{pages}{238101} (\bibinfo{year}{2013}).

\bibitem[{\citenamefont{Liebchen and Levis}(2017)}]{liebchen2017collective}
\bibinfo{author}{\bibfnamefont{B.}~\bibnamefont{Liebchen}} \bibnamefont{and}
  \bibinfo{author}{\bibfnamefont{D.}~\bibnamefont{Levis}},
  \bibinfo{journal}{Physical review letters} \textbf{\bibinfo{volume}{119}},
  \bibinfo{pages}{058002} (\bibinfo{year}{2017}).

\bibitem[{\citenamefont{Nagai et~al.}(2015)\citenamefont{Nagai, Sumino,
  Montagne, Aranson, and Chat{\'e}}}]{nagai2015collective}
\bibinfo{author}{\bibfnamefont{K.~H.} \bibnamefont{Nagai}},
  \bibinfo{author}{\bibfnamefont{Y.}~\bibnamefont{Sumino}},
  \bibinfo{author}{\bibfnamefont{R.}~\bibnamefont{Montagne}},
  \bibinfo{author}{\bibfnamefont{I.~S.} \bibnamefont{Aranson}},
  \bibnamefont{and}
  \bibinfo{author}{\bibfnamefont{H.}~\bibnamefont{Chat{\'e}}},
  \bibinfo{journal}{Physical review letters} \textbf{\bibinfo{volume}{114}},
  \bibinfo{pages}{168001} (\bibinfo{year}{2015}).

\bibitem[{\citenamefont{Holubec et~al.}(2021)\citenamefont{Holubec, Geiss,
  Loos, Kroy, and Cichos}}]{holubec2021finite}
\bibinfo{author}{\bibfnamefont{V.}~\bibnamefont{Holubec}},
  \bibinfo{author}{\bibfnamefont{D.}~\bibnamefont{Geiss}},
  \bibinfo{author}{\bibfnamefont{S.~A.} \bibnamefont{Loos}},
  \bibinfo{author}{\bibfnamefont{K.}~\bibnamefont{Kroy}}, \bibnamefont{and}
  \bibinfo{author}{\bibfnamefont{F.}~\bibnamefont{Cichos}},
  \bibinfo{journal}{Physical review letters} \textbf{\bibinfo{volume}{127}},
  \bibinfo{pages}{258001} (\bibinfo{year}{2021}).

\bibitem[{\citenamefont{Morin et~al.}(2015)\citenamefont{Morin, Caussin, Eloy,
  and Bartolo}}]{morin2015collective}
\bibinfo{author}{\bibfnamefont{A.}~\bibnamefont{Morin}},
  \bibinfo{author}{\bibfnamefont{J.-B.} \bibnamefont{Caussin}},
  \bibinfo{author}{\bibfnamefont{C.}~\bibnamefont{Eloy}}, \bibnamefont{and}
  \bibinfo{author}{\bibfnamefont{D.}~\bibnamefont{Bartolo}},
  \bibinfo{journal}{Physical Review E} \textbf{\bibinfo{volume}{91}},
  \bibinfo{pages}{012134} (\bibinfo{year}{2015}).

\bibitem[{\citenamefont{Costanzo}(2019)}]{costanzo2019milling}
\bibinfo{author}{\bibfnamefont{A.}~\bibnamefont{Costanzo}},
  \bibinfo{journal}{Europhysics Letters} \textbf{\bibinfo{volume}{125}},
  \bibinfo{pages}{20008} (\bibinfo{year}{2019}).

\bibitem[{\citenamefont{Gulich et~al.}(2018)\citenamefont{Gulich, Baglietto,
  and Rozenfeld}}]{gulich2018temporal}
\bibinfo{author}{\bibfnamefont{D.}~\bibnamefont{Gulich}},
  \bibinfo{author}{\bibfnamefont{G.}~\bibnamefont{Baglietto}},
  \bibnamefont{and} \bibinfo{author}{\bibfnamefont{A.~F.}
  \bibnamefont{Rozenfeld}}, \bibinfo{journal}{Physica A: Statistical Mechanics
  and its Applications} \textbf{\bibinfo{volume}{502}}, \bibinfo{pages}{590}
  (\bibinfo{year}{2018}).

\bibitem[{\citenamefont{Roy et~al.}(2019)\citenamefont{Roy, Shirazi, Jantzen,
  and Abaid}}]{roy2019effect}
\bibinfo{author}{\bibfnamefont{S.}~\bibnamefont{Roy}},
  \bibinfo{author}{\bibfnamefont{M.~J.} \bibnamefont{Shirazi}},
  \bibinfo{author}{\bibfnamefont{B.}~\bibnamefont{Jantzen}}, \bibnamefont{and}
  \bibinfo{author}{\bibfnamefont{N.}~\bibnamefont{Abaid}},
  \bibinfo{journal}{Physical review E} \textbf{\bibinfo{volume}{100}},
  \bibinfo{pages}{062415} (\bibinfo{year}{2019}).

\bibitem[{\citenamefont{Wu et~al.}(2021)\citenamefont{Wu, Li, Chen, Li, and
  Zhang}}]{wu2021pattern}
\bibinfo{author}{\bibfnamefont{Y.}~\bibnamefont{Wu}},
  \bibinfo{author}{\bibfnamefont{J.}~\bibnamefont{Li}},
  \bibinfo{author}{\bibfnamefont{D.}~\bibnamefont{Chen}},
  \bibinfo{author}{\bibfnamefont{X.}~\bibnamefont{Li}}, \bibnamefont{and}
  \bibinfo{author}{\bibfnamefont{H.-T.} \bibnamefont{Zhang}},
  \bibinfo{journal}{Europhysics Letters} \textbf{\bibinfo{volume}{134}},
  \bibinfo{pages}{50004} (\bibinfo{year}{2021}).

\bibitem[{\citenamefont{Lu et~al.}(2022)\citenamefont{Lu, Zhang, and
  Qin}}]{lu2022improved}
\bibinfo{author}{\bibfnamefont{X.}~\bibnamefont{Lu}},
  \bibinfo{author}{\bibfnamefont{C.}~\bibnamefont{Zhang}}, \bibnamefont{and}
  \bibinfo{author}{\bibfnamefont{B.}~\bibnamefont{Qin}},
  \bibinfo{journal}{Physica A: Statistical Mechanics and its Applications}
  \textbf{\bibinfo{volume}{587}}, \bibinfo{pages}{126553}
  (\bibinfo{year}{2022}).

\bibitem[{\citenamefont{You et~al.}(2023)\citenamefont{You, Yang, Li, Du, and
  Wang}}]{you2023modified}
\bibinfo{author}{\bibfnamefont{F.}~\bibnamefont{You}},
  \bibinfo{author}{\bibfnamefont{H.-X.} \bibnamefont{Yang}},
  \bibinfo{author}{\bibfnamefont{Y.}~\bibnamefont{Li}},
  \bibinfo{author}{\bibfnamefont{W.}~\bibnamefont{Du}}, \bibnamefont{and}
  \bibinfo{author}{\bibfnamefont{G.}~\bibnamefont{Wang}},
  \bibinfo{journal}{Applied Mathematics and Computation}
  \textbf{\bibinfo{volume}{438}}, \bibinfo{pages}{127565}
  (\bibinfo{year}{2023}).

\bibitem[{\citenamefont{Lei et~al.}(2023)\citenamefont{Lei, Xiang, Duan, and
  Peng}}]{lei2023exploring}
\bibinfo{author}{\bibfnamefont{X.}~\bibnamefont{Lei}},
  \bibinfo{author}{\bibfnamefont{Y.}~\bibnamefont{Xiang}},
  \bibinfo{author}{\bibfnamefont{M.}~\bibnamefont{Duan}}, \bibnamefont{and}
  \bibinfo{author}{\bibfnamefont{X.}~\bibnamefont{Peng}},
  \bibinfo{journal}{Journal of the Royal Society Interface}
  \textbf{\bibinfo{volume}{20}}, \bibinfo{pages}{20230176}
  (\bibinfo{year}{2023}).

\bibitem[{\citenamefont{Liu et~al.}(2021)\citenamefont{Liu, Xiang, Chang, Yan,
  Zhou, and Tang}}]{liu2021hierarchical}
\bibinfo{author}{\bibfnamefont{X.}~\bibnamefont{Liu}},
  \bibinfo{author}{\bibfnamefont{X.}~\bibnamefont{Xiang}},
  \bibinfo{author}{\bibfnamefont{Y.}~\bibnamefont{Chang}},
  \bibinfo{author}{\bibfnamefont{C.}~\bibnamefont{Yan}},
  \bibinfo{author}{\bibfnamefont{H.}~\bibnamefont{Zhou}}, \bibnamefont{and}
  \bibinfo{author}{\bibfnamefont{D.}~\bibnamefont{Tang}},
  \bibinfo{journal}{Drones} \textbf{\bibinfo{volume}{5}}, \bibinfo{pages}{74}
  (\bibinfo{year}{2021}).

\bibitem[{\citenamefont{Gr{\'e}goire and Chat{\'e}}(2004)}]{gregoire2004onset}
\bibinfo{author}{\bibfnamefont{G.}~\bibnamefont{Gr{\'e}goire}}
  \bibnamefont{and}
  \bibinfo{author}{\bibfnamefont{H.}~\bibnamefont{Chat{\'e}}},
  \bibinfo{journal}{Physical review letters} \textbf{\bibinfo{volume}{92}},
  \bibinfo{pages}{025702} (\bibinfo{year}{2004}).

\bibitem[{\citenamefont{Nagy et~al.}(2007)\citenamefont{Nagy, Daruka, and
  Vicsek}}]{nagy2007new}
\bibinfo{author}{\bibfnamefont{M.}~\bibnamefont{Nagy}},
  \bibinfo{author}{\bibfnamefont{I.}~\bibnamefont{Daruka}}, \bibnamefont{and}
  \bibinfo{author}{\bibfnamefont{T.}~\bibnamefont{Vicsek}},
  \bibinfo{journal}{Physica A: Statistical Mechanics and its Applications}
  \textbf{\bibinfo{volume}{373}}, \bibinfo{pages}{445} (\bibinfo{year}{2007}).

\bibitem[{\citenamefont{Chat{\'e} et~al.}(2008)\citenamefont{Chat{\'e},
  Ginelli, Gr{\'e}goire, and Raynaud}}]{chate2008collective}
\bibinfo{author}{\bibfnamefont{H.}~\bibnamefont{Chat{\'e}}},
  \bibinfo{author}{\bibfnamefont{F.}~\bibnamefont{Ginelli}},
  \bibinfo{author}{\bibfnamefont{G.}~\bibnamefont{Gr{\'e}goire}},
  \bibnamefont{and} \bibinfo{author}{\bibfnamefont{F.}~\bibnamefont{Raynaud}},
  \bibinfo{journal}{Physical Review E—Statistical, Nonlinear, and Soft Matter
  Physics} \textbf{\bibinfo{volume}{77}}, \bibinfo{pages}{046113}
  (\bibinfo{year}{2008}).

\bibitem[{\citenamefont{Cavagna et~al.}(2021)\citenamefont{Cavagna, Chaikin,
  Levine, Martiniani, Puglisi, and Viale}}]{cavagna2021vicsek}
\bibinfo{author}{\bibfnamefont{A.}~\bibnamefont{Cavagna}},
  \bibinfo{author}{\bibfnamefont{P.~M.} \bibnamefont{Chaikin}},
  \bibinfo{author}{\bibfnamefont{D.}~\bibnamefont{Levine}},
  \bibinfo{author}{\bibfnamefont{S.}~\bibnamefont{Martiniani}},
  \bibinfo{author}{\bibfnamefont{A.}~\bibnamefont{Puglisi}}, \bibnamefont{and}
  \bibinfo{author}{\bibfnamefont{M.}~\bibnamefont{Viale}},
  \bibinfo{journal}{Physical Review E} \textbf{\bibinfo{volume}{103}},
  \bibinfo{pages}{062141} (\bibinfo{year}{2021}).

\bibitem[{\citenamefont{Wang et~al.}(2008)\citenamefont{Wang, Xu, and
  Wang}}]{wang2008potential}
\bibinfo{author}{\bibfnamefont{J.}~\bibnamefont{Wang}},
  \bibinfo{author}{\bibfnamefont{L.}~\bibnamefont{Xu}}, \bibnamefont{and}
  \bibinfo{author}{\bibfnamefont{E.}~\bibnamefont{Wang}},
  \bibinfo{journal}{Proceedings of the National Academy of Sciences}
  \textbf{\bibinfo{volume}{105}}, \bibinfo{pages}{12271}
  (\bibinfo{year}{2008}).

\bibitem[{\citenamefont{Zhu et~al.}(2024)\citenamefont{Zhu, Yang, Zhang, Wang,
  Fang, and Wang}}]{zhu2024uncovering}
\bibinfo{author}{\bibfnamefont{L.}~\bibnamefont{Zhu}},
  \bibinfo{author}{\bibfnamefont{S.}~\bibnamefont{Yang}},
  \bibinfo{author}{\bibfnamefont{K.}~\bibnamefont{Zhang}},
  \bibinfo{author}{\bibfnamefont{H.}~\bibnamefont{Wang}},
  \bibinfo{author}{\bibfnamefont{X.}~\bibnamefont{Fang}}, \bibnamefont{and}
  \bibinfo{author}{\bibfnamefont{J.}~\bibnamefont{Wang}},
  \bibinfo{journal}{Proceedings of the National Academy of Sciences}
  \textbf{\bibinfo{volume}{121}}, \bibinfo{pages}{e2401540121}
  (\bibinfo{year}{2024}).

\bibitem[{\citenamefont{Li and Wang}(2015)}]{li2015quantifying}
\bibinfo{author}{\bibfnamefont{C.}~\bibnamefont{Li}} \bibnamefont{and}
  \bibinfo{author}{\bibfnamefont{J.}~\bibnamefont{Wang}},
  \bibinfo{journal}{Cancer research} \textbf{\bibinfo{volume}{75}},
  \bibinfo{pages}{2607} (\bibinfo{year}{2015}).

\bibitem[{\citenamefont{Yan et~al.}(2013)\citenamefont{Yan, Zhao, Hu, Wang,
  Wang, and Wang}}]{yan2013nonequilibrium}
\bibinfo{author}{\bibfnamefont{H.}~\bibnamefont{Yan}},
  \bibinfo{author}{\bibfnamefont{L.}~\bibnamefont{Zhao}},
  \bibinfo{author}{\bibfnamefont{L.}~\bibnamefont{Hu}},
  \bibinfo{author}{\bibfnamefont{X.}~\bibnamefont{Wang}},
  \bibinfo{author}{\bibfnamefont{E.}~\bibnamefont{Wang}}, \bibnamefont{and}
  \bibinfo{author}{\bibfnamefont{J.}~\bibnamefont{Wang}},
  \bibinfo{journal}{Proceedings of the National Academy of Sciences}
  \textbf{\bibinfo{volume}{110}}, \bibinfo{pages}{E4185}
  (\bibinfo{year}{2013}).

\bibitem[{\citenamefont{Xu et~al.}(2021)\citenamefont{Xu, Patterson, Staver,
  Levin, and Wang}}]{xu2021unifying}
\bibinfo{author}{\bibfnamefont{L.}~\bibnamefont{Xu}},
  \bibinfo{author}{\bibfnamefont{D.}~\bibnamefont{Patterson}},
  \bibinfo{author}{\bibfnamefont{A.~C.} \bibnamefont{Staver}},
  \bibinfo{author}{\bibfnamefont{S.~A.} \bibnamefont{Levin}}, \bibnamefont{and}
  \bibinfo{author}{\bibfnamefont{J.}~\bibnamefont{Wang}},
  \bibinfo{journal}{Proceedings of the National Academy of Sciences}
  \textbf{\bibinfo{volume}{118}}, \bibinfo{pages}{e2103779118}
  (\bibinfo{year}{2021}).

\bibitem[{\citenamefont{Li and Wang}(2014)}]{li2014landscape}
\bibinfo{author}{\bibfnamefont{C.}~\bibnamefont{Li}} \bibnamefont{and}
  \bibinfo{author}{\bibfnamefont{J.}~\bibnamefont{Wang}},
  \bibinfo{journal}{Proceedings of the National Academy of Sciences}
  \textbf{\bibinfo{volume}{111}}, \bibinfo{pages}{14130}
  (\bibinfo{year}{2014}).

\bibitem[{\citenamefont{Fang et~al.}(2019)\citenamefont{Fang, Kruse, Lu, and
  Wang}}]{fang2019nonequilibrium}
\bibinfo{author}{\bibfnamefont{X.}~\bibnamefont{Fang}},
  \bibinfo{author}{\bibfnamefont{K.}~\bibnamefont{Kruse}},
  \bibinfo{author}{\bibfnamefont{T.}~\bibnamefont{Lu}}, \bibnamefont{and}
  \bibinfo{author}{\bibfnamefont{J.}~\bibnamefont{Wang}},
  \bibinfo{journal}{Reviews of Modern Physics} \textbf{\bibinfo{volume}{91}},
  \bibinfo{pages}{045004} (\bibinfo{year}{2019}).

\bibitem[{\citenamefont{Wang}(2015)}]{wang2015landscape}
\bibinfo{author}{\bibfnamefont{J.}~\bibnamefont{Wang}},
  \bibinfo{journal}{Advances in Physics} \textbf{\bibinfo{volume}{64}},
  \bibinfo{pages}{1} (\bibinfo{year}{2015}).

\bibitem[{\citenamefont{Wu and Wang}(2013)}]{wu2013landscape}
\bibinfo{author}{\bibfnamefont{W.}~\bibnamefont{Wu}} \bibnamefont{and}
  \bibinfo{author}{\bibfnamefont{J.}~\bibnamefont{Wang}}, \bibinfo{journal}{The
  Journal of Physical Chemistry B} \textbf{\bibinfo{volume}{117}},
  \bibinfo{pages}{12908} (\bibinfo{year}{2013}).

\bibitem[{\citenamefont{Su et~al.}(2024)\citenamefont{Su, Cao, Wang, Jiang, and
  Hou}}]{su2024dynamical}
\bibinfo{author}{\bibfnamefont{J.}~\bibnamefont{Su}},
  \bibinfo{author}{\bibfnamefont{Z.}~\bibnamefont{Cao}},
  \bibinfo{author}{\bibfnamefont{J.}~\bibnamefont{Wang}},
  \bibinfo{author}{\bibfnamefont{H.}~\bibnamefont{Jiang}}, \bibnamefont{and}
  \bibinfo{author}{\bibfnamefont{Z.}~\bibnamefont{Hou}}, \bibinfo{journal}{Cell
  Reports Physical Science} \textbf{\bibinfo{volume}{5}}
  (\bibinfo{year}{2024}).

\bibitem[{\citenamefont{Wang et~al.}(2010)\citenamefont{Wang, Zhang, and
  Wang}}]{wang2010kinetic}
\bibinfo{author}{\bibfnamefont{J.}~\bibnamefont{Wang}},
  \bibinfo{author}{\bibfnamefont{K.}~\bibnamefont{Zhang}}, \bibnamefont{and}
  \bibinfo{author}{\bibfnamefont{E.}~\bibnamefont{Wang}}, \bibinfo{journal}{The
  Journal of chemical physics} \textbf{\bibinfo{volume}{133}}
  (\bibinfo{year}{2010}).

\end{thebibliography}

\end{document}